\documentclass[review]{elsarticle}
\usepackage{lineno}
\usepackage[latin1]{inputenc}
\usepackage{amsmath,amsfonts,amssymb,makeidx,graphicx,enumerate,epsfig, epstopdf,tabularx,placeins,ragged2e,tikz,url,hanging,attachfile2}
\usepackage[english]{babel}
\usepackage{color}
\usepackage{booktabs}
\usepackage{breqn}
\usepackage{multirow}
\usepackage{float}
\usepackage{listings}
\usepackage{hyperref}
\usepackage{threeparttable}

\usepackage{chemformula}

\usepackage{xcolor}

\newcommand{\orcid}[1]{\href{https://orcid.org/#1}{\textcolor[HTML]{A6CE39}{\aiOrcid}}}
\DeclareMathOperator{\Tr}{Tr}

\usepackage{hyperref}
\hypersetup{
    unicode=false,          
     colorlinks=true,       
    linkcolor={blue!50!black!90},       
    citecolor={blue!50!black!90},        
    filecolor={blue!50!black!90} ,      
    urlcolor={blue!50!black!90}           
}

\modulolinenumbers[5]

\journal{Physica B: Condensed Matter}

\begin{document}

\begin{frontmatter}

\title{Thermodynamic Stable Site for Interstitial alloy (N or O) in bcc-Refractory Metals using Density Functional Theory}

\author[mymainaddress,physicsaddress]
{Henry Martin}

\author[mymainaddress,mathaddress]{Peter Amoako-Yirenkyi \corref{mycorrespondingauthor}}
\cortext[mycorrespondingauthor]{Peter Amoako-Yirenkyi}
\ead{hmartin@knust.edu.gh/amoakoyirenkyi@knust.edu.gh}


\author[physicsaddress]{Eric K. K. Abavare}

\address[mymainaddress]{Center for Scientific and Technical Computing, National Institute for Mathematical Sciences, Kumasi, Ghana}

\address[physicsaddress]{Department of Physics, Kwame Nkrumah University of Science and Technology, Kumasi, Ghana}

\address[mathaddress]{Department of Mathematics, Kwame Nkrumah University of Science and Technology, Kumasi, Ghana}

\begin{abstract}
Plasticity in body centered cubic (bcc) refractory metals are largely due to the stress tensor induced either by solute or thermal activation. 
The mechanism of the solute atom(s) residence causes instability in such metals. Earlier research have considered the mechanism of oxygen (O) or carbon (C) in tungsten (W), even though the major component of the environment is nitrogen (N). In this article, the density functional theory (DFT) was employed to investigate the thermodynamic stable site for an interstitial solute (N or O) in the bcc refractory metals (Mo and Nb) by calculating the equilibrium and structural parameters, dissolution energetics and volumetric strain. The dissolution mechanism of all the relaxed solid solution structures were predicted to be an exothermic reaction from the supersaturated cell to the low concentration (1.82 at.\%) except for Mo-N solid solution.
Convergence of volumetric strain was observed at the low concentration of the solute. At this point, the solid solution of Mo-N and Mo-O had a less measure of global stress (less distortion) at the octahedral (o) site while that of Nb-N and Nb-O were at the tetrahedral (t) site. This certainly shows why this two bcc-refractory metals in groups VB (Nb) and VIB (Mo) of the periodic table exhibit different deformation behaviours giving their difference in site preference stability.

\end{abstract}

\begin{keyword}
\texttt{Stable site} \sep Thermodynamics \sep Solid solution \sep bcc-refractory-interstitial \sep DFT  

\end{keyword}

\end{frontmatter}

%
%
\section{Introduction}
\begin{figure}[H]
\begin{center}
\includegraphics[width=0.9\textwidth,height=.15\textheight]{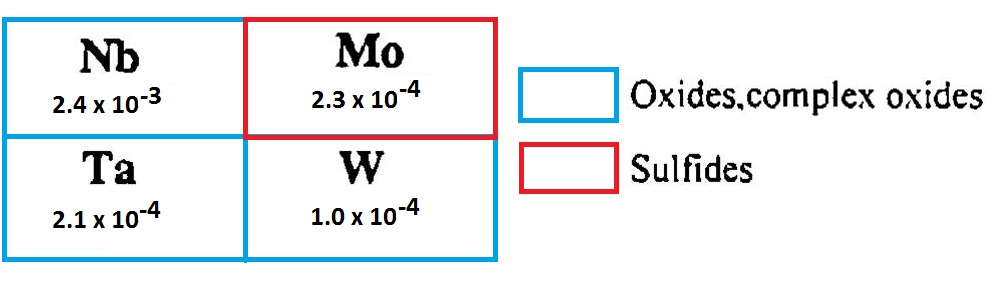}
\vspace{-0.52cm}
\caption{Natural forms of some refractory metals with relative percentage abundance aligned in groups VB and VIB of the periodic table  \citep{habashi2001historical}}
\label{Ref_Metals_Occurance_Form_Abandunce}
\end{center}
\end{figure}

Body centered cubic (bcc) refractory metals have extraordinary resistance to wear and heat giving their unusual high melting point \cite{davis2001alloying}. They also have high hardness value at room temperature and are chemically inert with relatively high density \cite{bauccio1993asm}. These metals are relatively abundant naturally in the earth's crust as mineral compounds of oxides and sulfides, Figure \ref{Ref_Metals_Occurance_Form_Abandunce} depicts few of such elements \cite{habashi2001historical}. The Niobium mineral: \emph{Pyrochlore} ($(Ca,Na)_{2-m} Nb_2O_6(O,OH,F)_{1-n} xH_2O$) and \emph{ Columbite}  ($(Fe,Mn)(Nb,Ta)_2 0_6$) are forms for obtaining its ore deposits.  
The only sulfide compound mineral is \emph{Molybdenite} ($MoS_2$), which is found as veins in quartz rock \citep{habashi2001historical}.

According to the Kepler conjecture, bcc crystal structures are less dense in comparison with fcc (ccp) and hcp \cite{Kepler1966,hales2017formal}. This, at low temperature warrants different behaviours in these structures during plastic deformation \cite{christian1983some,caillard2003thermally,argon2008strengthening}. Particularly at decreasing temperature, there exist rapid increase in the flow stress and strain rate of single crystals (thus, bcc metals) \cite{seeger1995flow,pichl2002slip,stukowski2015thermally,cereceda2016unraveling} which is mainly caused by breakdown in the standard geometric projection (non-glide components added to the stress tensor) \citep{chaussidon2006glide,brinckmann2008fundamental} during plastic deformation by solute \citep{penning1972mathematics,trinkle2005chemistry} or thermal activation \citep{romaner2010effect,li2012dislocation,samolyuk2012influence}.\\\
Many researchers have established that the bcc refractory metals of groups VB (V, Nb, and Ta) and VIB (Cr, Mo, and W) exhibit different deformation behaviours \citep{duesbery1998plastic,weinberger2013slip,weinberger2013peierls}. This difference is ascribed to the different behaviours of domination of screw dislocation motion at the cores during plastic deformation at low temperatures \citep{duesbery1998plastic,lim2013application,lim2018simulating}. 
Woodward and Rao demonstrated that the periodic table elements of group VB (Nb) has screw displacement spreading evenly about a central point while group VIB (Mo) are off centred \citep{woodward2001ab,dezerald2015first,dezerald2016plastic,rodney2017ab}.\\\ 
At lower temperatures, these bcc refractory metals display poor fracture toughness. Hence, the need to consider alloying with other elements to improve their ductility. A known technique called solid solution for producing novel materials has played significant role from the Bronze Age through the Iron Age to these advanced technological times \citep{hansen1958constitution,austin1978edgar}. This technique has so far caused a significant  change in the application of bcc metals strengthening (softening or hardening). 
\citep{martin2020statistical, martin2019, martin2002}. From the onset of this solid solution technique, several researches (experimentally and computationally) have sought to investigate and understand substitutional solid solution \citep{trinkle2005chemistry,medvedeva2007solid,romaner2010effect,zhao2011qm,li2012dislocation,samolyuk2012influence,itakura2013effect,romaner2014core,ventelon2015dislocation,hu2017solute,zhao2018direct}. Unfortunately, little attention has been given to the interstitial solid solution \citep{luthi2017first,svoboda2018anisotropy,zhao2019electronic}. 
\\\
However, some research has demonstrated that interstitial solutes such as B, C, N, O in bcc transition metal (Fe) stabilizes the hard core configuration of screw dislocation \citep{ventelon2015dislocation,rodney2017ab,luthi2018attractive,wang2019atomistic}. Recently, similar research has been done in bcc-refractory metal for C in (V, Nb, Ta, Mo and W) and O in W \citep{luthi2017first,zhao2019electronic}. These interstitial solutes have also been proven to produce a marked increase in the ductile-to-brittle transition temperature (DBTT) of bcc metals. This relation to DBTT effect is linked to the dislocation motion at a specific temperature level. A reduction of the kinetic energy (cause of low temperature) of the atoms slows the movement of dislocation in order for plasticity to take place. Typically, the addition of O and N to high pure crystals of Nb and Ta causes hardening \citep{stephens1964effects,ulitchny1973effects}. However, research show that the hardness of Nb decrease with increasing Nb concentration 
\citep{jiang1991elastic,jhi2001vacancy,chen2005hard,wu2005trends}. 
This shows the presence of the interstitial solutes retards the dislocation motion for the occurrence of plasticity. \\\ 
Given the importance of these bcc refractory interstitial solid solution in various applications where their plastic response ranges across length, time, and temperature scales. 
It is critical that we understand their performance and stability in any applicable regime on any scale \citep{hickman2004advanced,thakre2009chemical}.\\\ %
This paper addresses the thermodynamic mechanism occupancy preference of the interstitial solutes in bcc refractory metals. %
This was accomplished by using DFT because of the experimental difficulty of studying and measuring the effects of atomic distributions and external stresses on thermodynamic properties \citep{yan2018configurational}. The use of DFT offers an opportunity such as accessing the compositional temperature range of interest. This grants the ability to describe the deformation as well as to provide the essential chemical 
details of the interaction. Characteristics such as the equilibrium geometries, dissolution energetics and volumetric strain of the systems (MoN, NbN, NbO and MoO) were the center of focus for our investigation and analysis.
\section{Computational Method}
\begin{figure}[H]
\centering
\includegraphics[scale=0.8]{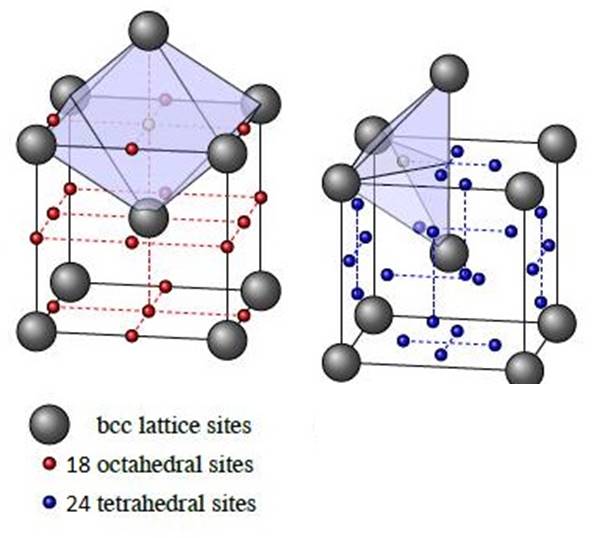}
\vspace*{-0.5cm}
\caption{Octahedral and tetrahedral sites of bcc unit cell \citep{zhao2019electronic}}\label{Meth:fig1}
\end{figure}

%
%
%
The Density Functional Theory (DFT) formulation \citep{hohenberg1964inhomogeneous,kohn1965self} as implemented in the Vienna ab initio Simulation Package (VASP) \citep{kresse1993software,kresse1996efficient,kresse1996efficiency} was employed. In employing DFT in the VASP package a construction of different sizes of supercell using a range of $1\times 1 \times 1 $ to $ 3 \times 3 \times 3$ was first done to depict different concentration of solute in the bcc refractory metal. This was done such that one interstitial solute (N or O) is inserted within an octahedral or tetrahedral sites of the bulk system (supercell of Mo or Nb) as shown in Figure \ref{Meth:fig1}.
The Projector Augmented Wave (PAW) pseudopotential scheme \citep{blochl1994projector,kresse1999ultrasoft} with semi-core p states was considered as part of the valence electrons for the bcc refractory metals \citep{dewaele2008compression,min2015study,liang_tel-01355132}.  The Perdew-Burke-Ernzerhof (PBE) functional \citep{perdew1996generalized,perdew1998nonlocality} within the Generalized Gradient Approximation (GGA) \citep{zupan1997distributions}, together with the Hermite-Gaussian within the Methfessel and Paxton method \citep{baldereschi1973mean,monkhorst1976special,methfessel1989high} were employed with an appropriate smearing sigma to obtain a negligible energy difference value less than 1 meV$/$atom.\\\
The convergence of kinetic energy cut-off was found with appropriate k-points 
leading to a stable thermodynamic system. For an accurate and validated comparison, the kinetic energy cut-off and the k-point chosen for a particular supercell structure (example Mo) is used for all systems except that of the molecules ($\ch{A2}= \ch{N2}$ or $\ch{O2}$). The systems used k-point sampling of $7\times 7 \times 7$, $3\times 3 \times 3$ and $3\times 3 \times 3$ 
for $X_2$, $X_{16}$ and $X_{54}$ 
respectively. Where, $X$ represent all systems. The molecules were computed by placing the dimer in a cubic box after which a $\Gamma$ point calculation was carried out.
%
%
The modified Broyden's method was applied for the final step of charge mixing scheme \citep{johnson1988modified}. The relaxation of both cell shape and atomic positions structures were performed until forces on each atom was less than $10^{-5}$ eV$/$atom using the Conjugate Gradient (CG) technique.\\\
The total energies of the interstitial solutes and the bulk refractory metals were obtained to calculate the dissolution energy  or the heat of solution at a specific x-site ($x=\{o$ or $t\}$), $E_{d_x}$ as
\begin{equation}
E_{d_x} = E_{\text{bcc cell} + \text{IS}} - E_{\text{bcc cell}} - \dfrac{1}{2} E[\ch{A2}]
\end{equation}
Where $E_{\text{bcc cell} + \text{IS}}$ and $E_{\text{bcc cell}}$ are the energies for bulk bcc refractory metal with and without an inserted interstitial solute respectively. The last term defines the energy of the molecular atom being considered. 
The dissolution energy difference, $\Delta E_{d_{o-t}}$, with respect to the difference in the occupied site ($x=\{o$ or $t\}$) was also obtained. Finally, the volume of the supercells were obtained after the relaxation of the whole system with and without the interstitial solute. This was done to calculate the volumetric strain which is equivalent to hydrostatic strain as follows:
\begin{equation}
\epsilon_v = \dfrac{\Delta V}{V} \approx \Tr(\epsilon) = 3 \epsilon
\end{equation}   
\section{Results and Discussion}
\begin{figure}[H]
\centering
\includegraphics[width=0.8\linewidth,height=.32\textheight]{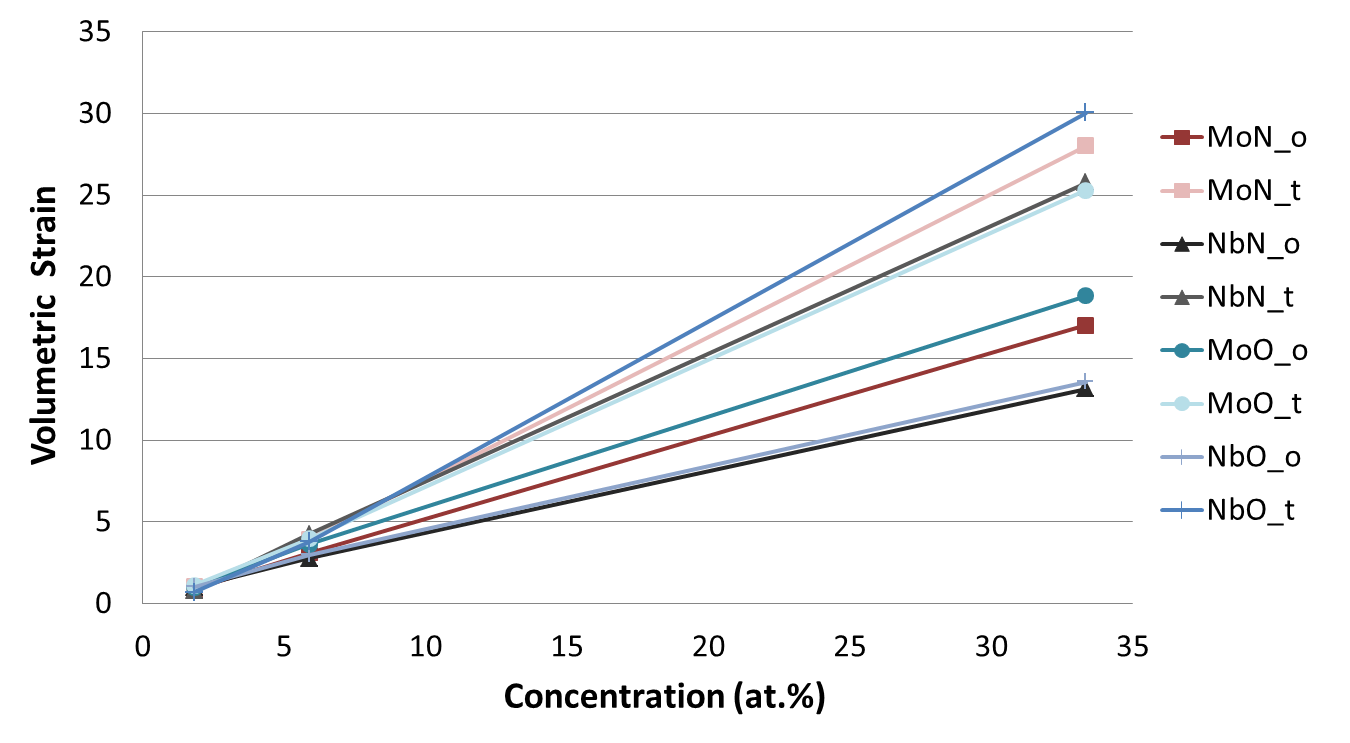}
\vspace*{-0.5cm}
\caption{A plot of volumetric strains when an interstitial solute (N or O) is inserted in the Octahedral (o) or Tetrahedral (t) site of a bcc refractory metal (Mo or Nb) at 0 K and relaxed 
}\label{RD:fig1}
\end{figure}

In determining the thermodynamic stable site of bcc refractory interstitial solid solution using the DFT technique, the equilibrium geometries were first obtained. The lattice constant of the bulk bcc refractory metals (Mo and Nb) and the vibration frequency of the dimers ($N_2$ and $O_2$) are shown in Table \ref{RD:tab1}. These as shown in the same Tables are within the range of experimental data as well as previously calculated DFT results. 
\begin{table}[H]
\caption{PAW-DFT-GGA predictions of lattice  parameters (a) of bcc refractory metals (Mo and Nb) and the vibration frequency ($cm^{-1}$) of the dimers (\ch{N2} and \ch{O2} gas) together with some other computational and experimental results are also shown for comparison}\label{RD:tab1}
\vspace*{-0.22cm}
\begin{center}
\begin{threeparttable}
  \begin{tabular}{ c  c | c  c  || c  c c c c}  \hline \hline
    Specie & a (\AA) &  Specie & a (\AA) &   Dimers & & Freq ($cm^{-1}$)  \\ \hline \hline
\multirow{6}{*}{Mo} & $3.0996^a$ & \multirow{6}{*}{Nb} & $3.294^g$ &  \multirow{3}{*}{$N_{2 \ {\tiny (gas)}}$} &  &  $2359^m$ & \\
  &  $3.142 \pm 0.03^b$  &    &$3.296^h$ &  &     &  $2566.28^n$ & \\
  & $3.15^{c,d}$  &  & $3.300^{i,j}$ &    &    & \textcolor{blue}{$\mathbf{2579.48}^q$}  & \\ \cline{5-8}
  &  $3.16^e$ &   & $3.312^k$  &   \multirow{3}{*}{$O_{2 \ {\tiny (gas)}}$}& & $1556^o$ & \\
  &    $3.17^f$ &   & $3.326^l$  &  &  & \textcolor{blue}{$\mathbf{1645.1}^q$} & \\
  &  \textcolor{blue}{$\mathbf{3.17}^q$} &  &\textcolor{blue}{$\mathbf{3.327}^q$} &   &   &  $2359^p$ &\\
 \hline
  \end{tabular}
  \begin{tablenotes}[para, flushleft]
      \footnotesize
      \item[$a$]Ref.\cite{trinkle2005chemistry}
      \item[$b$]Ref.\cite{davey1925precision}
      \item[$c$]Ref.\cite{kittel1996introduction}
      \item[$d$]Ref.\cite{hermann2016crystallography}
      \item[$e$]Ref.\cite{isaev2007phonon}
      \item[$f$]Ref.\cite{min2015study}
      \item[$g$]Ref.\cite{kittel1996introduction}
      \item[$h$]Ref.\cite{gray1972american}
      \item[$i$]Ref.\cite{delheusy2008x}
      \item[$j$]Ref.\cite{amriou2003fp}
      \item[$k$]Ref.\cite{haas2009calculation}
      \item[$l$]Ref.\cite{connetable2016first}
      \item[$m$]Ref.\cite{shimanouchi1972tables}
      \item[$n$]Ref.\cite{Nitrogen}
      \item[$o$]Ref.\cite{weber1960raman}
      \item[$p$]Ref.\citep{mcgrath1954influence}
      \item[$q$]This work
   \end{tablenotes}
  \end{threeparttable}
\end{center}
\end{table}
%
%
%
%
%
%
As displayed in Figure \ref{Meth:fig1}, 
the experimental lattice parameter ($a$) and atomic radius ($r$) of the bcc refractory metals (Mo is $3.15$ \AA \ and $1.363$ \AA \ and Nb is $3.30$ \AA \ and $1.430$ \AA \ respectively) were replaced in a theoretical formulae given as: 
\begin{equation}
\frac{r_o}{r} = \dfrac{a ( \frac{1}{2} - \frac{\sqrt{3}}{4})}{a \frac{\sqrt{3} }{4}} \tag{o-site}\label{RD:o-site}
\end{equation}
\begin{equation}
\dfrac{r_t}{r} = \dfrac{a (\frac{\sqrt{5}}{4}  - \frac{\sqrt{3}}{4})}{a \frac{\sqrt{3}}{4}} \tag{t-site}\label{RD:t-site}
\end{equation}
to obtain the radii of the o-site ($r_o$) and t-site ($r_t$) for Mo as $0.21086$ \AA \ and $0.39663$ \AA, respectively and for Nb as $0.22122$ \AA \ and $0.41612$ \AA, respectively. N and O has a covalent radius of $0.74$ \AA \ and $0.66$ \AA \ respectively. Therefore, the expectation would be that N and O will fit better in the t-site with less distortion, given that the radius at the t-site is wider than that of the o-site for fixed atoms position or Unrelaxed structure. We note that this theoretical expectation do not account for any form of relaxation within the geometry of the bcc lattice.\\\
In confirming this assertion from the theoretical formulae, we first assess Table \ref{RD:tab3} - \ref{RD:tab6} showing the dissolution energy difference of N and O occupancy in the o-site and t-site of bcc refractory metals supercell sizes using different concentration of N and O. Since the difference in dissolution energy determines the minimum energy at a particular site, the stability is gained for the alloying element located in the pure metal (bcc-refractory interstitial solid solution). It can be expressed as: 
\begin{equation}
\Delta E_{d_{o-t}} = + \ / \ -
\end{equation}
The ($+$) energy value $\Rightarrow$ $E_{d_o} > E_{d_t}$;  $\therefore$ t-site has \textcolor{blue}{minimum} energy while the ($-$) energy value $\Rightarrow$ $E_{d_o} < E_{d_t}$; $\therefore$ o-site has \textcolor{blue}{minimum} energy. This confirms the results of the theoretical formulae by considering the unrelaxed dissolution energy difference of all the bcc-refractory interstitial solid solution from Table \ref{RD:tab3} - \ref{RD:tab6} as well as the dissolution energies at the Appdx. \ref{APP:tab1} - \ref{APP:tab4}. Unfortunately, the relaxed structure was not permitted to conform to the same circumstance under which the theoretical formulae was set up. This is clearly shown for all concentration of the supercells or atleast either the 1st and 2nd supercells for the solid solutions. 
\begin{table}[H]
\begin{center}
\caption{Dissolution energy difference ($\Delta E_{d_{o-t}}$) between Nitrogen (N) in the o-site and t-site of bcc Molybdenum (Mo) for both unrelaxed (Ur) and relaxed (R) structures}\label{RD:tab3}
  \begin{tabular}{ c  c c c} \hline \hline
    Supercell & $C_N (at.\%)$ & $\Delta E_{{Ur}_{o-t}}$ (eV) & $\Delta E_{R_{o-t}}$ (eV)  \\ \hline \hline
    Mo2N1 & 33.3 & 2.0274 & -1.4906   \\ 
    Mo16N1 & 5.88 & 1.8335 & -0.5429  \\
    Mo54N1 &  1.82 & 2.0287 & -0.6960  \\
    \hline
  \end{tabular}
\end{center}
\end{table}
Starting with the concentration of $X_2 A$ cells which may be considered as a supersaturated solution of the interstitial in the bcc refractory metals. Its noticed that there was very high change in the dissolution energy towards the negative as compared to the other concentrations. This depicts stability at the octahedral sites which is in agreement with Table \ref{RD:tab7} and Figure \ref{RD:fig1} showing less volumetric strain at the o-site than that of the t-sites.
\begin{table}[H]
\begin{center}
\caption{Dissolution energy difference ($\Delta E_{d_{o-t}}$) between Nitrogen (N) in the o-site and t-site of bcc Niobium (Nb) for both unrelaxed (Ur) and relaxed (R) structures}\label{RD:tab4}
  \begin{tabular}{ c  c c c} \hline \hline
    Supercell & $C_N (at.\%)$ & $\Delta E_{{Ur}_{o-t}}$ (eV) & $\Delta E_{R_{o-t}}$ (eV)  \\ \hline \hline
    Nb2N1 & 33.3 & 1.3852 & -1.6695  \\ 
    Nb16N1 & 5.88  & 1.2986 & -0.9776 \\
    Nb54N1 &  1.82 & 1.4313 & 0.0041 \\
    \hline
  \end{tabular}
\end{center}
\end{table}
\begin{table}[H]
\begin{center}
\caption{Dissolution energy difference ($\Delta E_{d_{o-t}}$) between Oxygen (O) in the o-site and t-site of bcc Molybdenum (Mo) for both unrelaxed (Ur) and relaxed (R) structures}\label{RD:tab5}
  \begin{tabular}{ c  c c c} \hline \hline
    Supercell & $C_O (at.\%)$ & $\Delta E_{{Ur}_{o-t}}$ (eV) & $\Delta E_{R_{o-t}}$ (eV)  \\ \hline \hline
    Mo2O1 & 33.3 & 1.7413 & -0.5892   \\ 
    Mo16O1 & 5.88 & 2.7161 & 0.3629  \\
    Mo54O1 &  1.82 & 2.8144 & 0.0893  \\
    \hline
  \end{tabular}
\end{center}
\end{table}
\begin{table}[H]
\begin{center}
\caption{Dissolution energy difference ($\Delta E_{d_{o-t}}$) between Oxygen (O) in the o-site and t-site of bcc Niobium (Nb) for both unrelaxed (Ur) and relaxed (R) structures}\label{RD:tab6}
  \begin{tabular}{ c  c c c} \hline \hline
    Supercell & $C_O (at.\%)$ & $\Delta E_{{Ur}_{o-t}}$ (eV) & $\Delta E_{R_{o-t}}$ (eV)  \\ \hline \hline
    Nb2O1 & 33.3 & 1.4803 & -0.1599  \\ 
    Nb16O1 & 5.88  & 1.4312 & -0.8473 \\
    Nb54O1 &  1.82 & 1.6656 & 0.0336 \\
    \hline
  \end{tabular}
\end{center}
\end{table}
Unlike the unrelaxed structures, there is a change with the preference of the site occupancy when the concentration diminishes. This change of site preference from the o-site to the t-site in the concentration of $X_{16} A$ supercell is ONLY recognised in Mo16O1 (Table \ref{RD:tab5}) which is clearly seen among the other bcc refractory interstitial solid solution. It also shows a clear agreement with the volumetric strain shown in Table \ref{RD:tab7} even though the Mo16O1 depicts otherwise. This dispute with the dissolution energy difference and the volumetric strain, may be due to the slight change in shift or distortion as seen in Table \ref{RD:tab7}, which show that the volumetric strain at the o-site and t-site are quite closer as compared to other solid solution.
\begin{table}[H]
\begin{center}
\caption{Volumetric strain, $\epsilon_v = (\frac{\Delta V}{V})^*$  for the relaxed (R) cells structure of the interstitial solute, ($A = \{ N$ or $ O \}$) in the Octahedral (o) or Tetrahedral (t) site of the bcc refractory metal ($X = \{Mo$ or $Nb \}$)}\label{RD:tab7}
\vspace*{-0.2cm}
\scriptsize
\begin{threeparttable}
  \begin{tabular}{ c | c c | c c | c c | c c} \hline \hline
  \multirow{2}{*}{$C_A (at.\%)$}  & \multicolumn{2}{c}{${Mo}_x N1$}  & \multicolumn{2}{c}{${Nb}_x N1$}  & \multicolumn{2}{c}{${Mo}_x O1$} & \multicolumn{2}{c}{${Nb}_x O1$}  \\ \cline{2-9} \cline{2-9}
  & $\epsilon_{v_o}$ & $\epsilon_{v_t}$& $\epsilon_{v_o}$& $\epsilon_{v_t}$ & $\epsilon_{v_o}$  & $\epsilon_{v_t}$ & $\epsilon_{v_o}$  & $\epsilon_{v_t}$ \\ \hline
    33.3 & 17.0426 & 28.0075 & 13.1502 & 25.7466 & 18.8283 & 25.3133 & 13.5845 & 30.0358 \\ 
   5.88  & 3.1029  & 3.9317  &2.7804  & 4.2476 & 3.6135 & 3.9709 & 2.9754 & 3.7825 \\
   1.82 &  0.9074 & 1.0610 & 1.0028 & 0.8207 &  1.0063 & 1.1168 & 1.0160 & 0.6871 \\  \hline
  \end{tabular}
  \begin{tablenotes}[para, flushleft]
      \footnotesize
      \item[$*$] Volume increase of the relaxed $X_x A$ cell with respect to the relaxed $X_x$ cell.
   \end{tablenotes}
  \end{threeparttable}
\end{center}
\end{table}
The last supercell shows a complete change to the original site preference proposed by the theoretical formulae (thus, Eqs. \ref{RD:o-site} and \ref{RD:t-site}) as discussed earlier for all the solid solution except for Mo-N. This complete change again can be linked to the volumetric strain which is in total agreement with the assertion by taking a look at Figure \ref{RD:fig1} but has a slight contrast as shown in Table \ref{RD:tab7} with respect to the solid solution Mo-O. This slight contrast between the dissolution energy difference and the volumetric strain, is due to the closeness of the distortion that occur at both sites within the Mo54O1. Figure \ref{RD:fig1} shows the form of convergence with respect to the distortion and stability when the site preferences are of a concern to all solid solution. 
\section{Conclusion}
Density Functional Theory calculation was performed to examine the thermodynamic stable site preference on some selected bcc refractory interstitial solid solution. The dissolution mechanism of all the relaxed solid solution structures were predicted to be an exothermic reaction from the supersaturated cell to the low concentration (1.82 at.\%) except for Mo-N solid solution. Convergence of volumetric strain was observed moving towards low concentration (1.82 at.\%) of the solute with a less measure of global stress (less distortion) at the octahedral (o) site for the solid solutions of Mo-N and Mo-O and at the tetrahedral site for the Nb-N and Nb-O. This certainly shows why the two (2) bcc-refractory metals in groups VB (Nb) and VIB (Mo) of the periodic table exhibit different deformation behaviours giving their difference in site preference stability. We also noted that the demonstration of stability for all the solid solution at less concentration requires further study on the movement of the solute (N or O) across the sites (thus through o-sites, t-sites and o-t-sites) with a very low concentration of the solute or bigger supercell than what was under study in this work.
\section{Acknowledgements}
Centre for Scientific and Technical Computing, National Institute for Mathematical Sciences hosted by Kwame Nkrumah University of Science and Technology Kumasi, Ghana. The authors would like to thank Prof. Jaime Marian and his group at Department of Materials Science and Engineering and Mechanical and Aerospace Engineering, University of California Los Angeles (UCLA), Los Angeles, USA, for his helpful discussion,  insightful advice, access and usage of the computational and storage services associated with the Hoffman2 Shared Cluster provided by UCLA Institute for Digital Research and Education's Research Technology Group.
%
%
\section{Appendix}
\setcounter{table}{0}
\renewcommand{\thetable}{A\arabic{table}}
\begin{table}[H]
\begin{center}
\caption{Dissolution energy ($E_d$) of Nitrogen (N) in the Octahedral (o) and Tetrahedral (t) sites of bcc Molybdenum (Mo) for both unrelaxed (Ur) and relaxed (R) structures}\label{APP:tab1}
\begin{threeparttable}
  \begin{tabular}{ c  c | c c | c c} \hline \hline
\multirow{2}{*}{Supercell} & \multirow{2}{*}{$C_N (at.\%)$}  & \multicolumn{2}{c|}{o site}  & \multicolumn{2}{c}{t site}   \\ \cline{3-6} \cline{3-6} 
     &   & $ E_{d_{Ur}}$ (eV) & $ E_{d_R}$ (eV) & $ E_{d_{Ur}}$ (eV)  & $ E_{d_R}$ (eV) \\ \hline \hline
    Mo2N1 & 33.3 & 2.9713 & -0.6064  & 0.9439 & 0.8842 \\ 
    Mo16N1  & 5.88  & 5.4796  & 0.9791  & 3.6461 & 1.5220  \\
    Mo54N1 &  1.82 & 6.0527 & 0.7795 & 4.0240 & 1.4755  \\
    \hline
  \end{tabular}
  \end{threeparttable}
\end{center}
\end{table}
\begin{table}[H]
\begin{center}
\caption{Dissolution energy ($E_d$) of Nitrogen (N) in the Octahedral (o) and Tetrahedral (t) sites of bcc Niobium (Nb) for both unrelaxed (Ur) and relaxed (R) structures}\label{APP:tab2}
\begin{threeparttable}
  \begin{tabular}{ c  c | c  c | c c } \hline \hline
   \multirow{2}{*}{Supercell} & \multirow{2}{*}{$C_N (at.\%)$}  & \multicolumn{2}{c|}{o site}  & \multicolumn{2}{c}{t site}   \\ \cline{3-6} \cline{3-6} 
     &   & $ E_{d_{Ur}}$ (eV) & $ E_{d_R}$ (eV) & $ E_{d_{Ur}}$ (eV)  & $ E_{d_R}$ (eV) \\ \hline \hline
    Nb2N1& 33.3 & 1.0690 & -2.1949 & -0.3162 & -0.5253  \\ 
    Nb16N1  & 5.88 & 2.0490  & -1.7504 & 0.7504 & -0.7728  \\
    Nb54N1 &  1.82 & 2.2927 & -2.2332  & 0.8614  & -2.2373 \\ \hline
  \end{tabular}
  \end{threeparttable}
\end{center}
\end{table}
\vspace*{-0.2cm}
\begin{table}[H]
\begin{center}
\caption{Dissolution energy ($E_d$) of Oxygen (O) in the Octahedral (o) and Tetrahedral (t) sites of bcc Molybdenum (Mo) for both unrelaxed (Ur) and relaxed (R) structures}\label{APP:tab3}
\begin{threeparttable}
  \begin{tabular}{ c  c | c c | c c } \hline \hline
   \multirow{2}{*}{Supercell} & \multirow{2}{*}{$C_N (at.\%)$}  & \multicolumn{2}{c|}{o site}  & \multicolumn{2}{c}{t site}   \\ \cline{3-6} \cline{3-6} 
     &   & $ E_{d_{Ur}}$ (eV) & $ E_{d_R}$ (eV) & $ E_{d_{Ur}}$ (eV)  & $ E_{d_R}$ (eV) \\ \hline \hline
    Mo2O1 & 33.3 & 1.0479 & -1.4403 & -0.6934 & -0.8512 \\ 
    Mo16O1  & 5.88  & 4.7204 & -0.0226 & 2.0043 & -0.3855 \\
    Mo54O1 &  1.82 & 5.3028 & -0.3552 & 2.4885 & -0.4444 \\ \hline
  \end{tabular}
  \end{threeparttable}
\end{center}
\end{table}
\begin{table}[H]
\begin{center}
\caption{Dissolution energy ($E_d$) of Oxygen (O) in the Octahedral (o) and Tetrahedral (t) sites of bcc Niobium (Nb) for both unrelaxed (Ur) and relaxed (R) structures}\label{APP:tab4}
\begin{threeparttable}
  \begin{tabular}{ c  c | c c | c c } \hline \hline
   \multirow{2}{*}{Supercell} & \multirow{2}{*}{$C_N (at.\%)$}  & \multicolumn{2}{c|}{o site}  & \multicolumn{2}{c}{t site}   \\ \cline{3-6} \cline{3-6} 
     &   & $ E_{d_{Ur}}$ (eV) & $ E_{d_R}$ (eV) & $ E_{d_{Ur}}$ (eV)  & $ E_{d_R}$ (eV) \\ \hline \hline
    Nb2O1 & 33.3 & -1.1301 & -3.5987 & -2.6103 & -3.4388 \\ 
    Nb16O1  & 5.88  & 0.3354 & -3.6867 & -1.0958 & -2.8394 \\
    Nb54O1 &  1.82 & 0.6030 & -4.0744 & -1.0627 & -4.1080 \\ \hline
  \end{tabular}
  \begin{tablenotes}[para, flushleft]
      \footnotesize
      \item ($+$) pos energy value - endothermic process of dissolution.\\
      \item ($-$) neg energy value - exothermic process of dissolution
   \end{tablenotes}
  \end{threeparttable}
\end{center}
\end{table}
\newpage
%
%
%
%
\section*{References}
%
\bibliography{Bibliography}
\end{document}